\documentclass[prd,showpacs,nofootinbib]{revtex4}
\usepackage{amsmath}

\begin{document}
\title{Effects of chiral restoration on the behaviour of the
       Polyakov loop at strong coupling}
\author{Kenji Fukushima}
\email[E-mail: ]{fuku@nt.phys.s.u-tokyo.ac.jp}
\affiliation{Department of Physics, University of Tokyo, 7-3-1 Hongo,
             Bunkyo-ku, Tokyo 113-0033, Japan}
\begin{abstract}
We discuss the relation between the Polyakov loop and the chiral order
parameter at finite temperature. For that purpose we analyse an
effective model proposed by Gocksch and Ogilvie, which is constructed
by the double expansion of strong coupling and large dimensionality.
We make improvements in dealing with the model and then obtain
plausible results for the behaviours of both the Polyakov loop and the
chiral scalar condensate. The pseudo-critical temperature read from
the Polyakov loop turns out to coincide exactly with that read from
the chiral scalar condensate. Within the model study based on the
strong coupling expansion, the coincidence of the pseudo-critical
temperatures results from the fact that the jump of the Polyakov loop
in the presence of dynamical quarks should signify the chiral phase
transition rather than the deconfinement transition.
\end{abstract}
\maketitle

\paragraph*{\bf Introduction}
Quantum Chromodynamics (QCD) is commonly accepted as the fundamental
theory of the strong interaction. It has been intensely argued how a
thermodynamic system described by QCD goes through the phase
transitions at sufficiently high temperature, namely, the colour
deconfinement transition and the chiral phase transition around the
critical temperature $\sim150\:\mathrm{MeV}$~\cite{mey96}.

The analytic study of the deconfinement transition was first offered
within the framework of the strong coupling expansion on the
lattice~\cite{pol78}. In the absence of dynamical quarks a pure
gluonic system has the centre symmetry and the Polyakov loop provides
a criterion of confinement~\cite{sve82}. For the purpose of looking
into the spontaneous breaking of the centre symmetry, the effective
action in terms of the Polyakov loop variables has been constructed by
the strong coupling expansion, as well as by the perturbative
calculation. It is found that the resultant effective action leads to
a second-order phase transition~\cite{pol82} for the SU(2) gauge
theory and a first-order phase transition~\cite{gro83} for the
SU($N\geq3$) gauge theories. These are consistent with the
anticipation from the point of view of the universality
classifications~\cite{sve82} as well as of the lattice
observations~\cite{mcl81}. Also from the analysis of the effective
action it has been shown that the perturbative vacuum of the pure
gluonic system becomes unstable beyond some critical coupling
strength~\cite{fuk00}.

When dynamical quarks in the fundamental representation are present in
a theory, the centre symmetry is broken explicitly and the Polyakov
loop is no longer regarded as an order parameter to characterise the
deconfinement transition~\cite{ban83}. Then, the chiral symmetry
associated with light quarks plays an important role in hadronic
properties. There are many effective approaches based on the chiral
symmetry, such as the linear-sigma model, the Nambu--Jona-Lasinio
model, the chiral random matrix model and so on~\cite{mey96}. In those
model studies, features of colour confinement are hardly taken into
account. The relation between the confinement and the chiral symmetry
breaking has been discussed theoretically in a variety of contexts
such as the helicity conservation~\cite{cas79}, the anomaly matching
condition~\cite{tho80} and so on~\cite{cal77}. The difficulty in
studying their interrelation lies in the fact that no proper criterion
of confinement is established so far, despite of great
efforts~\cite{bri83}.

In the meanwhile, it has been found in the lattice simulations that
the Polyakov loop behaves like an approximate order parameter even in
the presence of dynamical quarks. Then the jumps of the Polyakov loop
and the chiral condensate are observed at the same critical
point~\cite{kog83}. Precise analyses thereafter in which the
pseudo-critical points are defined by respective susceptibilities have
sustained this coincidence~\cite{kar94}. It is often said that the
deconfinement and the chiral restoration should take place
simultaneously, though the physical reason for the coincidence is
still obscure.

The author of Ref.~\cite{sat98} proposed the following simple
explanation for the simultaneous jumps; at low temperature where the
chiral symmetry is spontaneously broken, the explicit breaking of the
centre symmetry is suppressed by the heavier mass of constituent
quarks rather than the lighter mass of current quarks. Consequently
the expectation value of the Polyakov loop stays small in the confined
or chiral broken phase at low temperature. Once the constituent quark
mass drops off in the chiral symmetric phase at high temperature, the
expectation value of the Polyakov loop no longer receives such
suppression. It follows that the jump of the Polyakov loop signifies
not the deconfinement transition but the chiral phase transition.
The simultaneous jumps are observed simply because the behaviours of
both the Polyakov loop and the chiral condensate indicate a single
phase transition of the chiral restoration.

This scenario had been partially confirmed within an effective model
constructed by the double expansion of strong coupling and large
dimensionality on the lattice~\cite{goc85}. In Ref.~\cite{goc85}
Gocksch and Ogilvie found that the given model leads to the
deconfinement transition at $T_{\mathrm{d}}=175\;\text{MeV}$ for the
pure gluonic system and the chiral restoration at
$T_\chi=471\;\text{MeV}$. The authors also argued that the constituent
quark mass tends to suppress the magnitude of the Polyakov loop,
though the simultaneous jumps could not be reproduced there. Actually
in Gocksch and Ogilvie's results, the transition temperatures relevant
to the deconfinement and the chiral restoration are too different to
affect each others.

The purpose of this letter is to show that the jump of the Polyakov
loop should be attributed to the chiral phase transition rather than
the deconfinement transition. To that end, we reexamine the model
given by Ref.~\cite{goc85}, which we call the Gocksch-Ogilvie model
hereafter. We make improvements in phenomenological approximations
employed in the original work and then find the deconfinement
transition at $T_{\mathrm{d}}=208\;\text{MeV}$ for the pure gluonic
system and the chiral phase transition at $T_\chi=162\;\text{MeV}$ in
the absence of the Polyakov loop dynamics. As compared with the
empirical values~\cite{mey96}, these transition temperatures are more
reasonable than those originally derived in Ref.~\cite{goc85}.
Furthermore, we analyse the coupled dynamics of the chiral order
parameter and the Polyakov loop in a numerical way and demonstrate
that the pseudo-critical temperature in regard to the Polyakov loop
coincides with the pseudo-critical temperature of the chiral
restoration at $T_{\mathrm{c}}=187\:\mathrm{MeV}$.
\vspace{3mm}

\paragraph*{\bf Gocksch-Ogilvie model}
The effective action of the Gocksch-Ogilvie model is constructed by
the double expansion of strong coupling, where the confining property
is almost trivial, and large dimensionality, where the dynamical
chiral symmetry breaking is realised on the lattice~\cite{klu83}. Here
we present only the expressions from Ref.~\cite{goc85};
\begin{align}
 S_{\text{eff}}[L,\lambda] &=-J\sum_{\mathrm{n.n.}}\mathrm{Tr_c}
  L(\vec{n})\mathrm{Tr_c}L^\dagger(\vec{m}) +\frac{N_{\mathrm{c}}}{2}
  \sum_{m,n}\lambda(n)V(n,m)\lambda(m)\notag\\
 &\qquad\qquad -\frac{1}{2}\sum_{\vec{n}}\biggl\{N_{\mathrm{c}}\ln
  \bigl[\cosh(N_\tau E)\bigr]+\mathrm{Tr_c}\ln\Bigl[1+\frac{1}
  {2\cosh(N_\tau E)}(L+L^\dagger)\Bigr]\biggr\},
\label{eq:eff_action}
\end{align}
where $\lambda(n)$ is the meson field, the condensation of which is
responsible for the dynamical breaking of the chiral symmetry. Since
the model is based on the staggered formalism on the lattice, the
flavour contents of meson fields are resolved in the differing lattice
sites. $N_{\mathrm{c}}$ is the number of colours. $L(\vec{n})$ denotes
the Polyakov loop defined in the $d$-dimensional space-time by
\begin{equation}
 L(\vec{n})=\prod_{n_4=1}^{N_\tau}U_d(\vec{n},n_4)
\end{equation}
with $N_\tau$ lattice sites in the thermal (temporal) direction. The
Polyakov loop is a matrix in the colour-space and $\mathrm{Tr_c}$
represents the trace over the colour indices. The static energy $E$ of
quasi-quarks is given by
\begin{equation}
 E =\sinh^{-1}\biggl(\sqrt{\frac{d-1}{2}}\lambda+m\biggr)
\end{equation}
with the current quark mass denoted by $m$. In the above expression
$(d-1)$ is the number of spatial dimensions. The propagator $V(n,m)$
of meson fields is given by
\begin{equation}
 V(n,m)=\frac{1}{2(d-1)}\sum_{\hat{j}}\bigl(\delta_{n,m+\hat{j}}+
  \delta_{n,m-\hat{j}}\bigr),
\end{equation}
where $\hat{j}$ runs over only the spatial directions. The strength of
the nearest neighbour interaction, $J$, is determined as a function of
the temperature, in other words, the temperature is specified by $J$:
In the strong coupling limit ($J\ll1$) we can express the correlation
function of the Polyakov loops in twofold ways as
\begin{equation}
 \langle\mathrm{Tr_c}L^\dagger(\vec{n})\mathrm{Tr_c}L(\vec{m})
  \rangle\sim J^{|\vec{m}-\vec{n}|}\sim
  \mathrm{e}^{-\beta\sigma a|\vec{m}-\vec{n}|},
\end{equation}
where $\sigma$ is the string tension and $a$ is the lattice spacing.
Accordingly the temperature in the Gocksch-Ogilvie model is fixed as
\begin{equation}
 T=\frac{1}{\beta}=\frac{\sigma a}{-\ln J}.
\label{eq:temperature}
\end{equation}

It should be emphasised that the coefficient $1/2$ in front of the
last part of Eq.~(\ref{eq:eff_action}) was missing in the original
work by Gocksch and Ogilvie. This factor is important in counting
the number of flavours from the phenomenological point of view. But
for the coefficient $1/2$, the scalar field $\lambda(n)$ generates too
many mesons with four degenerate light flavours peculiar to the
staggered formalism. This coefficient effectively reduces the model to
that with only two ($u$ and $d$) light quarks, as often adopted in
lattice simulations.

The prescription of taking the square root of the Dirac determinant is
reliable only in the weak coupling limit where the Dirac operator is
well localised. In the present case, the prescription is also
reliable, though the model is based on the strong coupling expansion.
The counterpart of the Dirac operator is absolutely local because the
last part of Eq.~(\ref{eq:eff_action}) comes from the integration with
respect to the quark field without its kinetic term which is absorbed
in that of mesons.

With the factor $1/2$, the formulae at zero temperature given in
Ref.~\cite{klu83} should receive modifications. Certainly the
expectation value of $\lambda$ at zero temperature becomes
$\sqrt{1/2}$ from $1$, but all the physical quantities such as the
hadron masses and the chiral condensate are not affected if just the
current quark mass is divided by $\sqrt{1/2}$. Actually there are two
parameters inherent in the model, namely, the lattice spacing $a$ and
the current quark mass $m$. According to the formulae given in
Ref.~\cite{klu83} with modifications from the factor $1/2$, the model
parameters can be fixed as follows so as to reproduce the pion mass
$140\:\mathrm{MeV}$ and the $\rho$ meson mass $770\:\mathrm{MeV}$:
\begin{equation}
 a^{-1}=432\:\mathrm{MeV},\qquad m=5.7\:\mathrm{MeV}.
\label{eq:parameter}
\end{equation}

We would emphasise the following point: In the zero temperature
analysis like Ref.~\cite{klu83}, too many light flavours would cause
no serious problem as long as the mass spectrum in only the $u$ and
$d$ quark sector is concerned. In the chiral limit ($m=0$) the
mean-field approximation leads to the same answer for each quark
flavour regardless of the flavour number. In the presence of small
current quark mass, the difference of the flavour number can be
absorbed in the difference of the current quark mass. At finite
temperature, on the other hand, how many degrees of freedom are
excited at that temperature should affect significantly the relevant
temperature to the chiral restoration. In this case, since the number
of pions (Nambu-Goldstone bosons) is too large, the resultant
potential energy prevents the chiral restoration until considerably
high temperature is reached.\footnote{This might seem contradict to
the fact that the chiral transition temperature is lowered as the
number of flavours increases. In the treatment of the Gocksch-Ogilvie
model, however, we exploit the mean-field approximation and any
fluctuation of pions, which is responsible for reducing the transition
temperature, is frozen from the beginning. Thus the potential energy
alone is involved in the analysis and it gives higher transition
temperature with more flavours.} Thus the factor $1/2$ is important in
order to estimate the chiral transition temperature quantitatively.
\vspace{3mm}

\paragraph*{\bf Mean-field approximation}
Following the treatment of Ref.~\cite{goc85}, we adopt the mean-field
approximation to deal with the effective action~(\ref{eq:eff_action}).
As to the meson field, we simply replace it by a condensate
$\bar{\lambda}$. The Polyakov loop $L(\vec{n})$ is integrated out with
the mean-field action~\cite{cre83},
\begin{equation}
 S_{\text{mf}}[L]=-\frac{x}{2}\sum_{\vec{n}}\bigl\{\mathrm{Tr_c}
  L(\vec{n})+\mathrm{Tr_c}L^\dagger(\vec{n})\bigr\},
\label{eq:mf_action}
\end{equation}
where $x$ is the variational parameter to be determined afterwards
from the extremal condition on the free energy. The mean-field free
energy is then given by
\begin{align}
 \beta f_{\text{mf}}(x,\bar{\lambda}) &=-N^{-(d-1)}\Bigl(\langle
  -S_{\text{eff}}[L,\bar{\lambda}]+S_{\text{mf}}[L]\rangle_{\text{mf}}
  +\ln\int\mathcal{D}L\,\mathrm{e}^{-S_{\text{mf}}[L]}\Bigr)\notag\\
 &=\beta f_{\text{mf}}^{\mathrm{g}}(x)
  +\beta f_{\text{mf}}^{\mathrm{q}}(x,\bar{\lambda}),
\end{align}
where $\langle\cdots\rangle_{\text{mf}}$ stands for the expectation
value calculated by means of the mean-field
action~(\ref{eq:mf_action}). In the expression of the mean-field free
energy, the physical implication of each term is plainly understood as
follows: In the first line, the first term is the internal energy
tending to make the system ordered and the remaining terms can be
regarded as the entropy tending to make the system disordered.
$f_{\text{mf}}^{\mathrm{g}}(x)$ represents the pure gluonic part of
the mean-field free energy and
$f_{\text{mf}}^{\mathrm{q}}(x,\bar{\lambda})$ corresponds to the
chiral part with the coupling between the Polyakov loop and the scalar
condensate. They are explicitly written as
\begin{align}
 \beta f_{\text{mf}}^{\mathrm{g}}(x) &=-2(d-1)J\Bigl(\frac{\mathrm{d}}
  {\mathrm{d}x}\ln I(x)\Bigr)^2 +x^2\frac{\mathrm{d}}{\mathrm{d}x}
  \Bigl(\frac{1}{x}\ln I(x)\Bigr),
\label{eq:free_energy_g}\\
 \beta f_{\text{mf}}^{\mathrm{q}}(x,\bar{\lambda})
  &=\frac{N_{\mathrm{c}}N_\tau}{2}\bar{\lambda}^2
  -\frac{N_{\mathrm{c}}}{2}\ln\bigl[\cosh(N_\tau E)\bigr]
  -\frac{1}{2}\frac{\tilde{I}(x;\cosh(N_\tau E))}{I(x)},
\label{eq:free_energy_q}
\end{align}
where $I(x)$ is defined by~\cite{kog82}
\begin{equation}
 I(x)=\int\mathrm{d}L\,\exp\Bigl[\frac{x}{2}\mathrm{Tr_c}(L+
  L^\dagger)\Bigr] =N_{\mathrm{c}}!\sum_m\det I_{m-i+j}(x),
\label{eq:def_I}
\end{equation}
and $\tilde{I}(x;\alpha)$ is given by
\begin{align}
 \tilde{I}(x;\alpha)&=\sum_{a=1}^{N_{\mathrm{c}}}
  \sum_{m=-\infty}^\infty\epsilon_{i_1i_2\dots i_{N_{\mathrm{c}}}}
  \epsilon_{j_1j_2\dots j_{N_{\mathrm{c}}}}I_{m-i_1+j_1}(x)\dots
  \tilde{I}_{m-i_a+j_a}(x;\alpha)\dots I_{m-i_{N_{\mathrm{c}}}+
  j_{N_{\mathrm{c}}}}(x) \notag\\
 \tilde{I}_n(x;\alpha)&=\int_0^{2\pi}\frac{\mathrm{d}\phi}{2\pi}
  \ln\Bigl[1+\frac{\cos\phi}{\alpha}\Bigr]\,\mathrm{e}^{x
  \cos\phi+\mathrm{i}n\phi}.
\end{align}
Here $I_n(x)$ denotes the modified Bessel function of the first kind.
Once we expand $\tilde{I}_n(x;\alpha)$ in terms of $1/\alpha$ and
furthermore expand the above expressions in terms of $x$, we can
immediately reproduce the results of Ref.~\cite{goc85}. These
expansions in terms of $1/\alpha$ and $x$ are not always regarded as
reliable in fact. Although the expansion in terms of $x$ works well
for the estimate of $I(x)$, it is not a good approximation for
$f_{\mathrm{mf}}^{\mathrm{g}}(x)$ due to significant cancellations in
the right-hand-side of Eq.~(\ref{eq:free_energy_g}). The expansion in
terms of $1/\alpha$ becomes uncertain when the expectation value of
the Polyakov loop is large. Therefore we do not adopt such expansions
in examining the mean-field free energies in the present work.

The deconfinement transition in the pure gluonic system is described
by $f_{\mathrm{mf}}^{\mathrm{g}}(x)$, while
$f_{\mathrm{mf}}^{\mathrm{q}}(x,\bar{\lambda})$ determines the fate of
the chiral symmetry in the presence of the Polyakov loop. Actually
$f_{\mathrm{mf}}^{\mathrm{g}}(x)$ results in the first-order phase
transition at $2(d-1)J_{\mathrm{c}}=0.806$ in the case of
$N_{\mathrm{c}}=3$. The corresponding temperature read from
Eq.~(\ref{eq:temperature}) is $T_{\mathrm{d}}=208\:\mathrm{MeV}$ with
the numerical value of the string tension
$\sigma=(425\:\mathrm{MeV})^2$ substituted.

Then, we will look into the chiral dynamics somewhat closely in an
analytic way in the case of the chiral limit ($m=0$). When the phase
transition is second-order, that is the case in the present analysis,
the transition temperature can be deduced from the condition that the
quadratic term with respect to the order parameter changes its sign,
or passes across zero. Thus we can derive the analytic expression for
$T_\chi$ from the expansion of
$f_{\mathrm{mf}}^{\mathrm{q}}(x,\bar{\lambda})$ in terms of
$\bar{\lambda}$.

The calculation is simplified when the Polyakov loop is put as
trivial, \textit{i.e.}, $L\simeq 1$ for the moment. Then we can
approximate the last term of Eq.~(\ref{eq:free_energy_q}) by
$-(N_{\mathrm{c}}/2)\ln[1+1/\cosh(N_\tau E)]$ looking back at
Eq.~(\ref{eq:eff_action}), which leads to
\begin{equation}
 \beta f_{\mathrm{mf}}^{\mathrm{q}}(\bar{\lambda})\simeq
  \frac{N_{\mathrm{c}}N_\tau}{2}\bar{\lambda}^2-\frac{(d-1)
  N_{\mathrm{c}}N_\tau^2}{16}\bar{\lambda}^2+(\text{const.}).
\label{eq:expand_f}
\end{equation}
From this expansion, the chiral phase transition temperature in the
absence of the Polyakov loop dynamics is immediately determined as
\begin{equation}
 T_\chi =\frac{d-1}{8}a^{-1}=162\:\mathrm{MeV}
\label{eq:temperature_chi}
\end{equation}
with the numerical value of $a$ given by Eq.~(\ref{eq:parameter}).
The above expression is essentially the same as (3.23) in
Ref.~\cite{goc85}.

Another interesting limit in which the calculation is simplified is
that the Polyakov loop is forced to vanish, \textit{i.e.}, $x=0$. Then
it follows that
\begin{equation}
 \biggl\langle\mathrm{Tr_c}\ln\Bigl[1+\frac{1}{2\cosh(N_\tau E)}
  (L+L^\dagger)\Bigr]\biggr\rangle_{\text{mf}} \simeq -N_{\mathrm{c}}
  \ln2+\sqrt{d-1}N_{\mathrm{c}}N_\tau|\bar{\lambda}|-\frac{(d-1)
  N_{\mathrm{c}}N_\tau^2}{4}\bar{\lambda}^2.
\label{eq:linear}
\end{equation}
The quadratic term proportional to $\bar{\lambda}^2$ is exactly
cancelled by the contribution from $N_{\mathrm{c}}\ln[\cosh(N_\tau E)]
\simeq(d-1)N_{\mathrm{c}}N_\tau^2\bar{\lambda}^2/4$ and only the
tree-term $N_{\mathrm{c}}N_\tau\bar{\lambda}^2/2$ remains. What is
surprising is that the linear term of $\bar{\lambda}$ appears in
Eq.~(\ref{eq:linear}). Owing to the presence of this linear term, the
stationary point with respect to $\bar{\lambda}$ always leaves from
zero. In other words, \textit{the chiral symmetry is broken at any
temperature}. If the vanishing Polyakov loop, $x=0$, has something to
do with confinement even in the presence of dynamical quarks, this
result suggests that the chiral symmetry must be broken in the
confined vacuum, which is in agreement with the arguments in
Refs.~\cite{cas79,tho80}.

Hence, the chiral dynamics is substantially affected by the Polyakov
loop dynamics. It means at the same time that the behaviour of the
Polyakov loop is deeply related to that of the chiral order
parameter.
\vspace{3mm}

\paragraph*{\bf Numerical results}
We must search for the global minimum of the free energy
$f_{\mathrm{mf}}(x,\bar{\lambda})$ numerically in general, except for
the above special cases of $m=0$, $L\simeq1$ and $m=0$, $x=0$ where
analytic treatments are feasible. The variational parameter, $x$, and
the scalar condensate, $\bar{\lambda}$, are determined as functions
of temperature. Then we can readily acquire the expectation value of
the Polyakov loop by using (see Eq.~(\ref{eq:def_I}))
\begin{equation}
 \Bigl\langle\frac{1}{2N_{\mathrm{c}}}\bigl(\mathrm{Tr_c}L
  +\mathrm{Tr_c}L^\dagger\bigr)\Bigr\rangle=\frac{1}{N_{\mathrm{c}}}
  \frac{\mathrm{d}}{\mathrm{d}x}I(x).
\end{equation}

The results are shown in Fig.~\ref{fig:result}. It is apparent that
the behaviours like phase transitions are observed around the
temperature $T_{\mathrm{c}}\simeq180\:\mathrm{MeV}$ in
Fig.~\ref{fig:result}. As compared with the pure gluonic result
$T_{\mathrm{d}}=208\:\mathrm{MeV}$ (shown by the dotted curve for
reference) it seems that the critical point concerning the Polyakov
loop dynamics is smeared by dynamical quarks and, as a result, two
crossovers in terms of the Polyakov loop and the scalar condensate are
observed nearly at the same point.

To make the argument more quantitative, we can define the
pseudo-critical temperatures by means of the peak of susceptibilities
in a similar way to Ref.~\cite{kog83}. Here we make use of the
simplest ones, which are immediately feasible in the present analysis,
that is, the temperature susceptibilities
$\chi_{\mathrm{t}}^L=\partial\langle\mathrm{Tr_c}L\rangle/\partial
T/N_{\mathrm{c}}$ and $\chi_{\mathrm{t}}^\lambda=-\partial
\bar{\lambda}/\partial T$. The results are shown in
Fig.~\ref{fig:sus}. The pseudo-critical temperatures are found exactly
at the same point $T_{\mathrm{c}}=187\:\mathrm{MeV}$. As to the
moderate peak of $\chi_{\mathrm{t}}^\lambda$ slightly above
$\sim200\:\mathrm{MeV}$, it should be regarded as an accidental
artifact because this peak would vanish if the current quark mass $m$
is raised or lowered by hand.

The coincidence of the pseudo-critical temperatures completely agrees
with the lattice observations. Within the Gocksch-Ogilvie model, the
jump of the Polyakov loop does not mean the deconfinement transition.
Actually it is caused by the last term of Eq.~(\ref{eq:eff_action})
through which the behaviour of the Polyakov loop would reflect that of
the scalar condensate. In analogy with the hopping parameter expansion
(see Eq.~(2.11) in Ref.~\cite{gre84}) we can regard
$1/\cosh(N_\tau E)$ as the strength of an external field to break the
centre symmetry. The behaviour of the Polyakov loop is governed
effectively by the chiral dynamics through this strength of an
external field.

We consider that the exact coincidence of the pseudo-critical
temperatures in the present study provides a credible support for the
argument of Ref.~\cite{sat98}: The behaviour of the Polyakov loop
indicates only the chiral phase transition rather than the
deconfinement transition. Thus the coincidence of the critical
temperatures is only a consequence of viewing a single phenomenon,
\textit{i.e.}, the chiral restoration.

Finally we shall comment upon the plans for further work. Since the
number of the space-time dimensions is four at most, it is necessary
to make sure the convergence of the large dimensional expansion and to
confirm the quantitative success in describing the Polyakov loop and
the chiral dynamics even with next-to-leading order contributions
taken into account.

It would be intriguing to construct a similar effective model in the
formulation of the Wilson fermion~\cite{ros96}. Then the flavour
decomposition becomes straightforward at the cost of the intricate
structure of the Dirac indices. Also an extension to introduce adjoint
fermions, which does not break the centre symmetry, would be
interesting. In contrast to the case of fundamental fermions, two
distinct transitions are found in the lattice
simulations~\cite{kar99}. It would be challenging to reproduce such
results within the model study parallel to the present treatment. Such
analyses would serve as a double-check for the speculation on the
behaviour of the Polyakov loop discussed in this letter.

\begin{acknowledgments}
The author is supported by Research Fellowships of the Japan Society
for the Promotion of Science for Young Scientists. He thanks S.~Sasaki
for comments.
\end{acknowledgments}

\vspace{5mm}

\begin{figure}[ht]
\caption{The behaviours of the order parameters as functions of the
temperature. Transition-like jumps are observed simultaneously around
$T_{\mathrm{c}}\simeq180\:\mathrm{MeV}$. The dotted curve is the
result in the pure gluonic case for reference.}
\label{fig:result}

\caption{The temperature susceptibilities for the Polyakov loop and
the chiral scalar condensate.}
\label{fig:sus}
\end{figure}

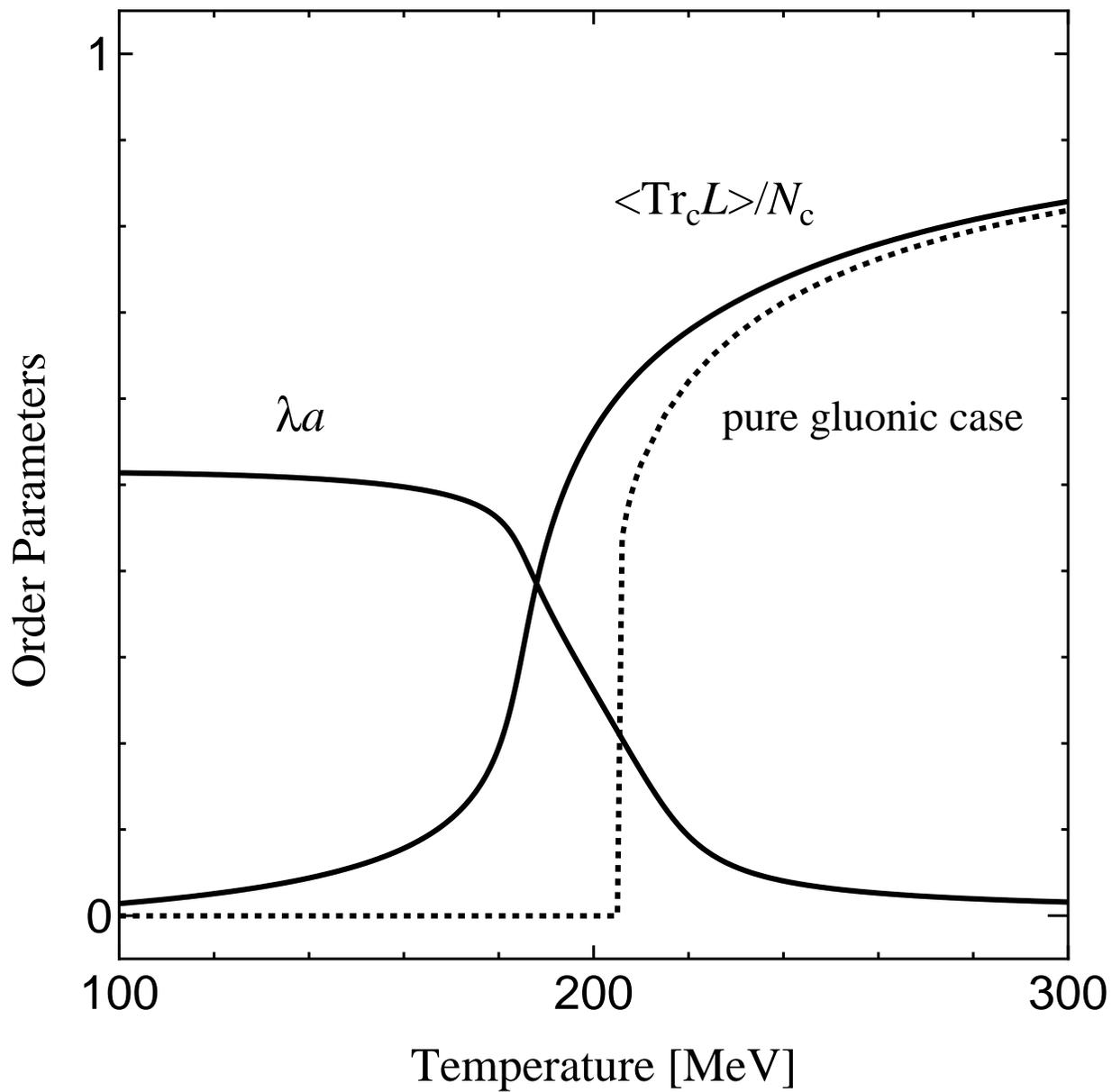
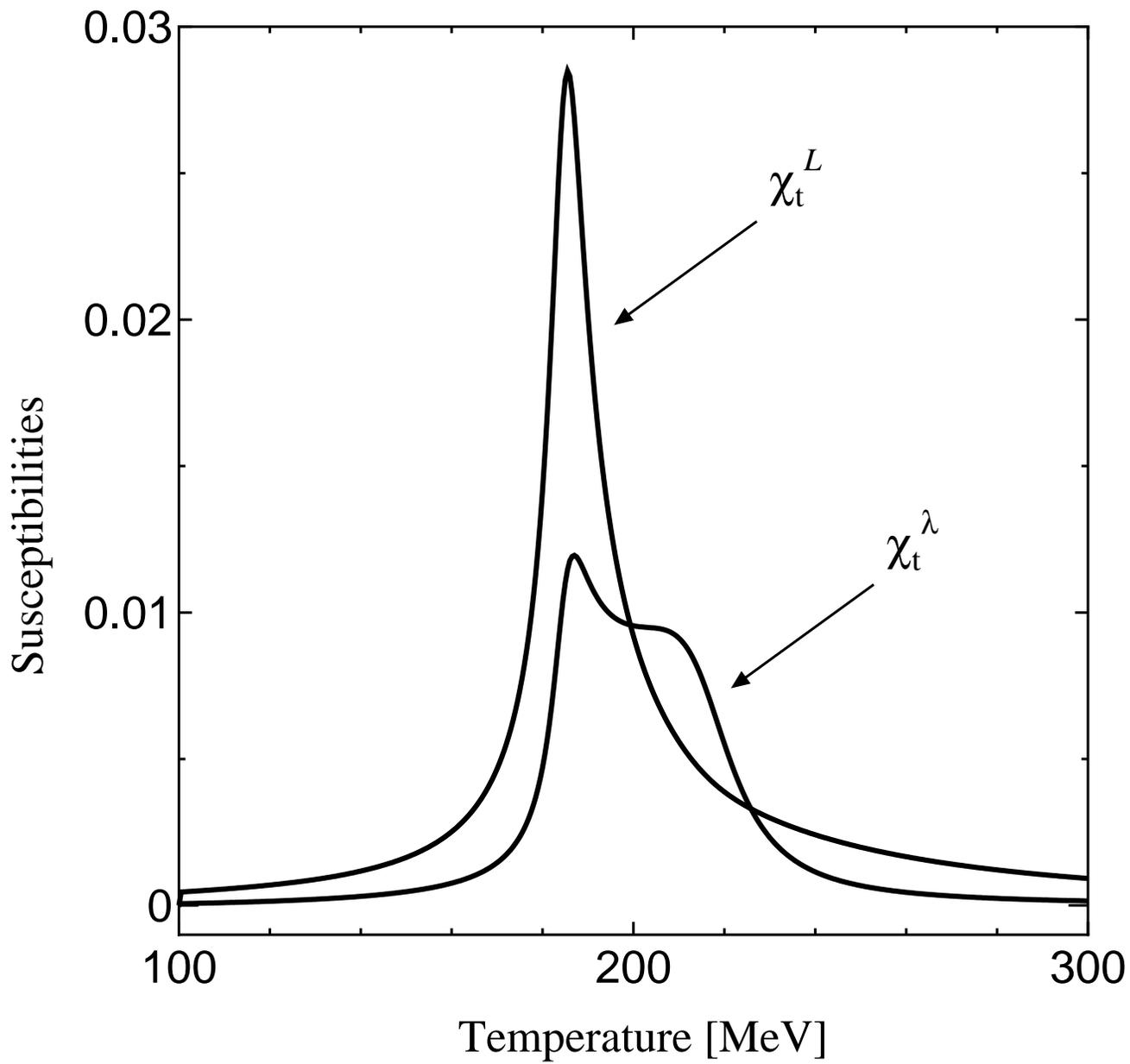
\end{document}